\documentclass[11pt]{article}
\usepackage{epsbox}
\begin{document}

\title{SU(3) Lattice QCD Study for Static Three-Quark Potential}

\author{
T. T. Takahashi$^1$, H. Suganuma$^2$, H. Matsufuru$^3$ and
Y. Nemoto$^3$\\ \\
$^1$RCNP, Osaka University, Mihogaoka 10-1, Osaka 567-0047, Japan\\
$^2$Faculty of Science, Tokyo Institute of Technology, Tokyo 152-8551, Japan\\
$^3$YITP, Kyoto University, Kitashirakawa, Sakyo, Kyoto 606-8502, Japan
}

\maketitle

\begin{abstract}
We study the static three-quark (3Q) potential in detail using SU(3) 
lattice QCD with $12^3 \times 24$ at $\beta=5.7$ and 
$16^3 \times 32$ at $\beta=5.8, 6.0$ at the quenched level. 
For more than 200 patterns of the 3Q systems, we numerically derive 
3Q ground-state potential $V_{\rm 3Q}$ 
from the 3Q Wilson loop with the smearing technique, 
which reduces excited-state contaminations. 
The lattice QCD data of $V_{\rm 3Q}$ are well reproduced 
within a few \% deviation by a sum of a constant, the two-body Coulomb term 
and the three-body linear confinement term $\sigma_{\rm 3Q} L_{\rm min}$, 
with $L_{\rm min}$ the minimal value of total length 
of color flux tubes linking the three quarks. 
From the comparison with the Q-$\bar {\rm Q}$ potential, 
we find a universality of the string tension as $\sigma_{\rm 3Q} \simeq 
\sigma_{\rm Q \bar Q}$ and the one-gluon-exchange result for Coulomb 
coefficients, $A_{\rm 3Q} \simeq \frac12 A_{\rm Q \bar Q}$. 
\end{abstract}

\maketitle

\section{Introduction}
Strong interaction in hadrons or nuclei is fundamentally 
governed by quantum chromodynamics (QCD).
In a spirit of the elementary particle physics, it would be desirable 
to construct hadron physics and nuclear physics at quark level based on QCD.
For instance, to bridge between QCD and hadron physics, 
the inter-quark potential\cite{TMNS01,TMNS99,SMNT01} is one of the most important quantities, 
because it is directly responsible for the hadron structure 
at the quark level\cite{CI86}.
However, it still remains as a difficult problem to derive 
the inter-quark potential analytically from QCD,  
because of the strong-coupling nature of QCD in the infrared region. 

Recently, the lattice QCD calculation has been adopted as a 
useful and reliable method for the nonperturbative analysis of QCD.
For instance, quark-antiquark (Q$\bar{\rm Q}$) potential, 
which is responsible for the meson properties, has been well studied 
using lattice QCD\cite{TMNS01,BSS95}.
%``Confinement'' is one of the difficult problems.
%Quark-antiquark(${\rm Q\bar{Q}}$) potential which is directly
%responsible for meson features which carries nuclear force has been
%well studied using lattice QCD calculations, which is a powerful tool
%to solve QCD directly.
\begin{figure}[h]
\begin{center}
%\resizebox{0.45\columnwidth}{!}{\includegraphics{qqbar60.eps}}
\centerline{\epsfile{file=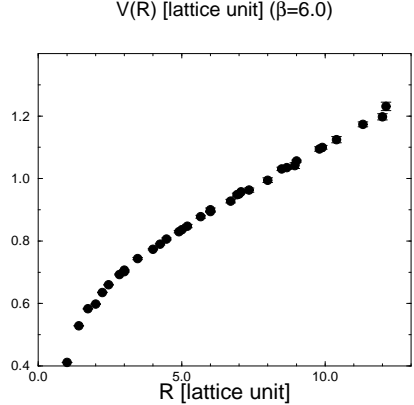,width=0.45\columnwidth}}
\caption{
The ${\rm Q\bar{Q}}$ static potential $V_{\rm Q\bar{Q}}(r)$ 
obtained in SU(3) lattice QCD at the quenched level. 
}
\label{qqbarpot}
\end{center}
\end{figure}
In Fig.1, we show the ${\rm Q\bar{Q}}$ static potential  
$V_{\rm Q\bar{Q}}(r)$ as the function of 
the distance $r$ between quark and antiquark 
in SU(3) lattice QCD at the quenched level. 
Here, the ${\rm Q\bar{Q}}$ potential can be well reproduced by a sum of the 
Coulomb term due to the perturbative one-gluon-exchange (OGE) process, 
a linear confinement term and a constant\cite{TMNS01,BSS95}, 
\begin{eqnarray}
V_{\rm Q\bar{Q}}=-\frac{A_{\rm Q\bar{Q}}}{r}+\sigma_{\rm Q\bar{Q}}r+
C_{\rm Q\bar{Q}},
\label{qqbar}
\end{eqnarray}
where the string tension $\sigma_{\rm Q\bar{Q}} \simeq $ 0.89 GeV/fm 
represents the confinement force. 
The linear potential at the long distance can be physically 
interpreted with the flux-tube picture or the string picture 
for hadrons\cite{CI86,N74,KS75,SST95}. 
In this picture, the quark and the antiquark are linked with a 
one-dimensional flux-tube with the string tension $\sigma_{\rm Q\bar{Q}}$. 
Hence, ${\rm Q\bar{Q}}$ potential in the long-range region is
proportional to the distance $r$ between quark and antiquark.
This flux-tube picture (or the string picture) in the infrared region 
is supported by the Regge trajectory of hadrons, 
empirical analysis of heavy quarkonia data, 
the strong-coupling expansion of QCD\cite{KS75}
and recent lattice QCD simulations.

However, there is almost no reliable formula to describe the three-quark (3Q)
potential $V_{\rm 3Q}$ directly based on QCD, 
besides the strong-coupling QCD\cite{CI86,KS75}, 
although $V_{\rm 3Q}$ is directly responsible for the baryon
properties\cite{CI86,BPV95,RS91}. Up to now, the 3Q potential has been treated 
phenomenologically or hypothetically more than 20 years.
%So we study 3Q potential using lattice QCD.
We carry out the detailed study of the 3Q potential using SU(3) lattice QCD.

\section{Theoretical Consideration}

Theoretically, the 3Q potential\cite{TMNS01,SMNT01,CI86} is 
expected to be expressed by a sum of a constant, 
the two-body Coulomb term from the perturbative OGE process at the short distance and  
the three-body linear confinement term at the long distance, 
similarly in the ${\rm Q \bar Q}$ potential. 
Here, reflecting the SU(3)$_c$ gauge theory in QCD, 
the color flux tube has a junction which combines 
three different colors, $R$, $B$ and $G$, in a color-singlet manner.
For most 3Q systems, the flux-tube energy is minimized 
with the presence of the junction, and therefore 
a Y-type flux tube is expected to be formed among the three quarks
\cite{TMNS01,CI86,KS75,BPV95,RS91}.
We show in Fig.2 the flux-tube configuration which minimizes 
the total flux-tube length in the 3Q system. 

Thus, the static 3Q potential is considered to take a form as  
\begin{eqnarray}
V_{\rm 3Q}=-A_{\rm 3Q}\sum_{i<j}\frac1{|{\bf r}_i-{\bf r}_j|}
+\sigma_{\rm 3Q} L_{\rm min}+C_{\rm 3Q}, 
\label{3Qp}
\end{eqnarray}
where $L_{\rm min}$ denotes the miminal value of the total length of  
color flux tubes linking three quarks. 

\begin{figure}[h]
\begin{center}
%\resizebox{0.27\columnwidth}{!}{\includegraphics{y2.eps}}
\centerline{\epsfile{width=0.27\columnwidth,file=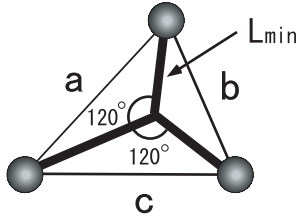}}
\caption{The flux-tube configuration of the 3Q system with 
the minimal value of the total flux-tube length $L_{\rm min}$. }
\label{lmin}
\end{center}
\end{figure}

Denoting by $a$, $b$ and $c$ the three side lengths of the 3Q triangle 
as shown in Fig.2, $L_{\rm min}$ is explicitly expressed as \cite{TMNS01} 
\begin{eqnarray}
L_{\rm min}=\left[\frac12 (a^2+b^2+c^2)
 +\frac{\sqrt{3}}2 
  \sqrt{(a+b+c)(-a+b+c)(a-b+c)(a+b-c)}\right]^{\frac12},
\end{eqnarray}
when any angle of the 3Q triangle does not exceed 2$\pi$/3.
For the case with one angle larger than 2$\pi$/3, $L_{\rm min}$ is given as
$
L_{\rm min}=(a+b+c)-\max (a,b,c).
$
%$\max (a,b,c)$ takes the largest value of $a$,$b$ and $c$.

\section{Static 3Q Potential in SU(3) Lattice QCD}
\begin{figure}[h]
\begin{center}
%\resizebox{0.25\columnwidth}{!}{\includegraphics{Bloop.staple.eps}}
\centerline{\epsfile{width=0.25\columnwidth,file=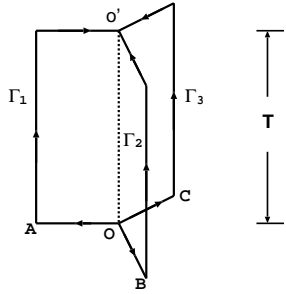}}
\caption{The 3Q Wilson loop in the Euclidean space-time. 
$\Gamma_k (k=1,2,3)$ denotes the path linking $O$ and $O'$.
}
\label{3Ql}
\end{center}
\end{figure}

The static 3Q potential can be extracted from the 3Q Wilson loop \cite{TMNS01,BPV95,RS91}
in SU(3) lattice QCD calculations.
As shown in Fig.\ref{3Ql}, the 3Q Wilson loop physically represents 
the creation of a gauge-invariant 3Q state at $t=0$,  
the system with spatially fixed three quarks in $0 < t < T$ and 
the annihilation of the 3Q state at $t=T$. 

In four-dimensional Euclidean space-time, 
the 3Q Wilson loop is a color current defined 
in a gauge-invariant manner as 
\begin{eqnarray}
W_{\rm 3Q} \equiv \frac1{3!}\varepsilon_{abc}\varepsilon_{a'b'c'}
U_1^{aa'} U_2^{bb'} U_3^{cc'}, \quad 
U_k=P \exp \{ i g \int_{\Gamma_k} dx^\mu A_\mu(x) \} \in {\rm SU(3)}.  
\label{3qope}
\end{eqnarray}
Here, $P$ denotes the path-ordered product along the path denoted by 
${\Gamma_k} (k=1,2,3)$ linking $O$ and $O'$ in Fig.\ref{3Ql}.
Similar to the derivation of the Q${\bar {\rm Q}}$ potential 
from the Wilson loop, the 3Q potential $V_{\rm 3Q}$ can be obtained as 
$V_{\rm 3Q}=-\lim_{T \rightarrow \infty} \frac1T 
\ln \langle W_{\rm 3Q}\rangle$.

Next, let us consider the physical state of the 3Q system. 
The ground state of the 3Q system is expected to be composed 
by flux tubes rather than the strings\cite{TMNS01,SST95}, and we denote by 
$|{\rm g.s.;} t \rangle$ the 3Q ground state at $t$. 
On the other hand, there are many excited states of the 3Q system 
corresponding to the flux-vibrational modes\cite{TMNS01}, and we denote by 
$|n{\rm th e.s.;} t \rangle$ the $n$-th excited 3Q state at $t$.
In the 3Q Wilson loop, 
the gauge-invariant 3Q state created at $t=0$ 
and annihilated at $t=T$ can be expressed as
\begin{eqnarray}
&&|{\rm 3Q}; 0\rangle =c_0|{\rm g.s.;0}\rangle+c_1|{\rm 1st \ e.s.;0}\rangle+
c_2|{\rm 2nd \ e.s.;0}\rangle+...,\\ \nonumber
&&|{\rm 3Q}; T\rangle =c_0|{\rm g.s.;} T \rangle+c_1|{\rm 1st \ e.s.;} T \rangle+
c_2|{\rm 2nd \ e.s.;} T \rangle+...,
\end{eqnarray}
where the coefficients are normalized as $\sum_{i=0}^\infty |c_i|^2=1$.
(This normalization is found to be consistent 
with the definition of $W_{\rm 3Q}$ in Eq.(\ref{3qope}).) 
Then, the expectation value of $W_{\rm 3Q}$ can be expressed as
\begin{eqnarray}
\langle W_{\rm 3Q}\rangle&=&
|c_0|^2\langle{\rm g.s.;}T|{\rm g.s.;0}\rangle+
|c_1|^2\langle{\rm 1st \ e.s.;}T|{\rm 1st \ e.s.;0}\rangle+...\\ \nonumber
&=& |c_0|^2\exp(-V_{\rm g.s.}T)+|c_1|^2\exp(-V_{\rm 1st\ e.s.}T)+... \, 
\label{expan}
\end{eqnarray}
with the ground-state potential $V_{\rm g.s.}$ of the 3Q system and 
the $n$-th excited-state potential $V_{n\hbox{-}{\rm th\ e.s.}}$.

As increasing $T$, the excited-state components drop faster than the 
ground-state component in $\langle W_{\rm 3Q} \rangle$, however, the 
ground-state component $|c_0|^2\exp(-V_{\rm g.s.}T)$ also 
decreases exponentially.
Hence, we face a practical difficulty in extracting the numerical signal.
To avoid this difficulty, we adopt the smearing technique\cite{TMNS01,BSS95,APE87}
which enhances the ground-state overlap and removes the excited-state contamination efficiently.
(There were a few lattice studies on the 3Q potential more than 
13 years ago\cite{SW8486,TES88}, however, their results seem unreliable and misleading 
due to the fatal large excited-state contaminations.  
The smearing technique was mainly developed after their works.)

\section{Smearing Technique to Remove Excited Modes}
\begin{figure}[h]
\begin{center}
%\resizebox{0.6\columnwidth}{!}{\includegraphics{smearing.eps}}
\centerline{\epsfile{width=0.6\columnwidth,file=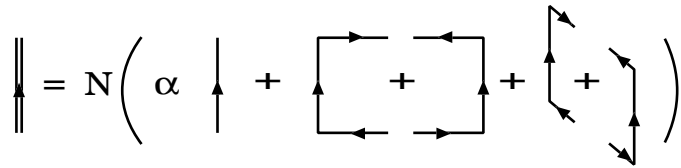}}
\caption{The schematic explanation of the smearing for the link-variables.}
\label{smr}
\end{center}
\end{figure}
The smearing for the link-variable is expressed as the iterative replacement of the spatial 
link-variables by a obscured link-variables $\bar U_i(s) \in {\rm SU(3)}$ which maximizes 
\begin{eqnarray}
{\rm Re} \,\,{\rm tr} \left\{ \bar U_i^{\dagger}(s) \left[
\alpha U_i(s)+\sum_{j \ne i} \{
U_j(s)U_i(s+\hat j)U_j^\dagger (s+\hat i) + 
U_j^\dagger (s-\hat j)U_i(s-\hat j)U_j(s+\hat i-\hat j)
\} \right] \right\}
\end{eqnarray}
with a real smearing parameter $\alpha$.
This can be visualized as in Fig.\ref{smr}.
%We show this process visually in Fig.\ref{smr}.
%In Fig.\ref{smr} ``N'' denotes the projection operator to SU(3) matrix.
The obscured spatial line composed by the obscured link-variables 
physically corresponds to the flux tube with a finite cylindrical radius. 
The smearing parameter $\alpha$ and the iteration number of the smearing are suitably 
determined so as to maximize the ground-state overlap of the 3Q system 
at $t=0$ and $t=T$ in the 3Q Wilson loop. 
In fact, after a suitable smearing, we expect saturation of the ground-state overlap
as $|c_0|^2 \simeq 1$, i.e. strong reduction of the excited-state contaminations as 
$|c_i|^2\simeq 0$ ($i=1,2,..$)\cite{TMNS01,BSS95,APE87}. 
\begin{figure}[h]
\begin{center}
\centerline{\epsfile{height=4.8cm,file=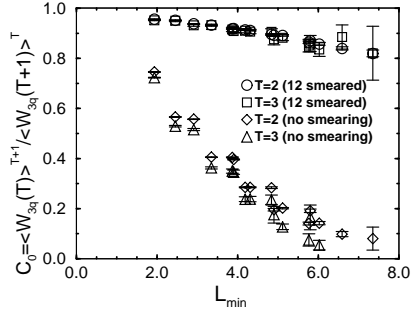}}
\caption{The ground-state overlap $C_0 \equiv 
\langle W_{\rm 3Q}(T)\rangle^{T+1}/\langle W_{\rm 3Q}(T+1)\rangle^{T}$ in SU(3) lattice QCD at $\beta=5.7$.
By the smearing, the ground-state overlap $C_0$ is largely enhanced from 
the lower data to the upper data as $0.8 < C_0 < 1$ for each 3Q system.}
\label{czero}
\end{center}
\end{figure}
To see this in actual lattice QCD calculations, we examine in Fig.\ref{czero} the ground-state overlap 
\begin{eqnarray}
C_0 \equiv \langle W_{\rm 3Q} (T) \rangle ^{T+1} / 
\langle W_{\rm 3Q} (T+1) \rangle ^T 
\end{eqnarray}
corresponding to $|c_0|^2$\cite{TMNS01,BSS95}, 
and we observe a large enhancement of the ground-state overlap as $0.8 < C_0 < 1$ by the smearing.
To summarize here, owing to the smearing, we can set up 
the ground-state dominant 3Q state at $t=0$ and $t=T$ in the 3Q Wilson loop, and therefore 
we enjoy accurate measurements for the 3Q potential  
without suffering from excited modes like the flux-tube vibration.

\section{The Lattice QCD Results for The 3Q Potential}

\vspace{0.5cm}
\begin{figure}[h]
\begin{center}
%\resizebox{0.4\columnwidth}{!}{\includegraphics{3Qloop_57_nor.eps}}
%\resizebox{0.4\columnwidth}{!}{\includegraphics{3Qloop_58_nor.eps}}
\centerline{\epsfile{width=0.4\columnwidth,file=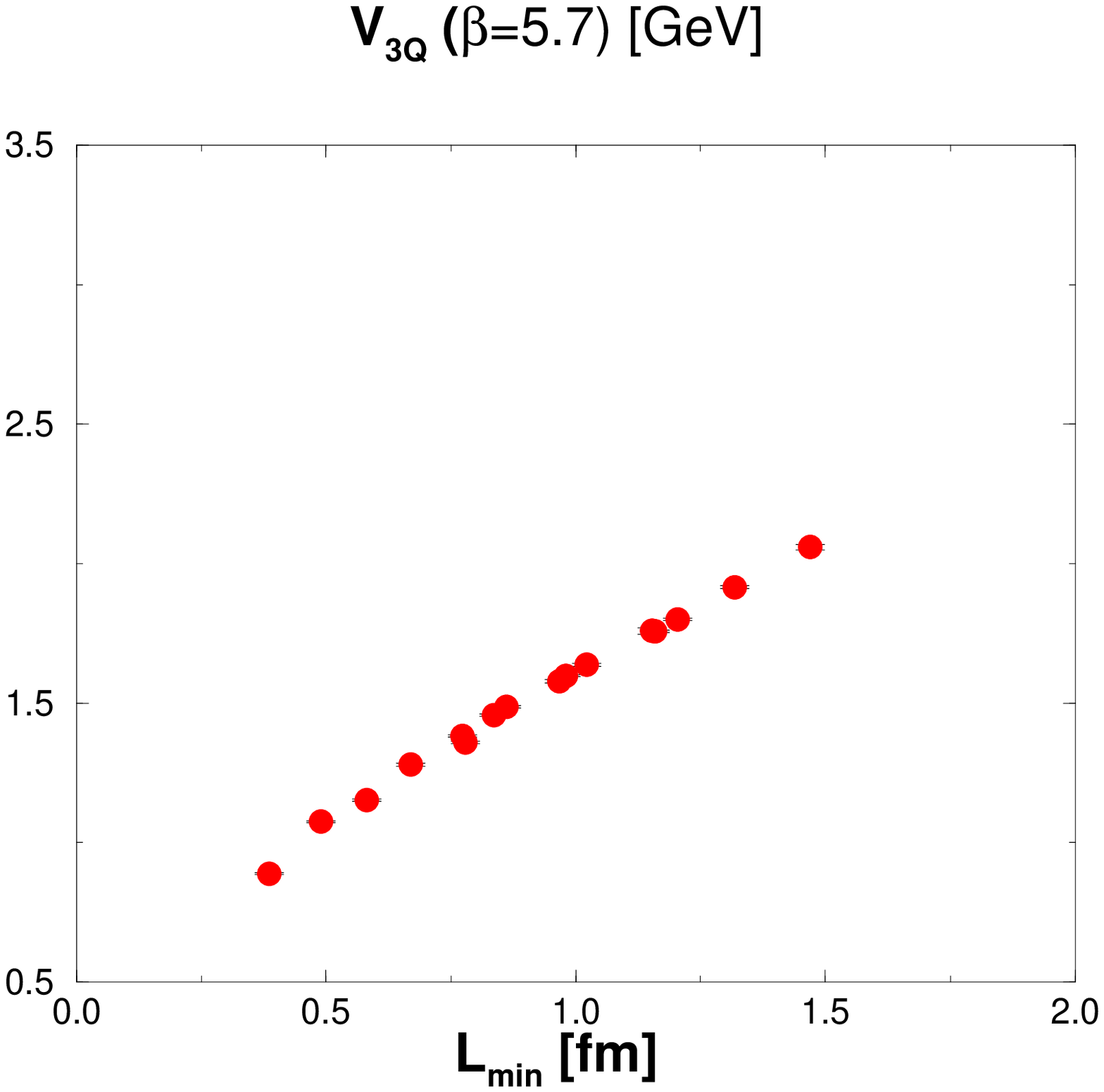}
\epsfile{width=0.4\columnwidth,file=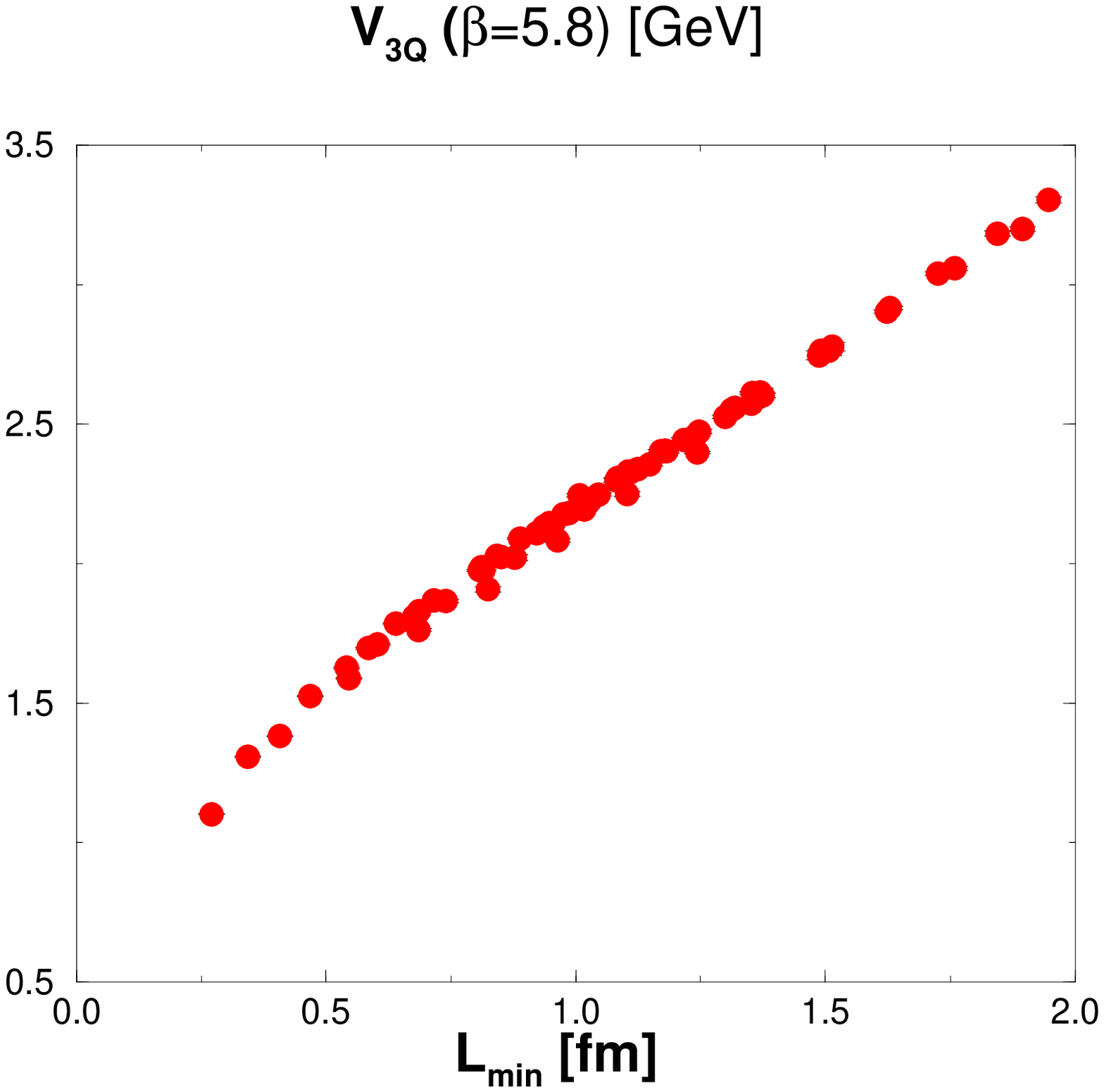}}
\caption{The 3Q potential $V_{\rm 3Q}$ as the function of $L_{\rm min}$, 
the minimal value of the total flux-tube length,  
in SU(3) lattice QCD at $\beta=5.7$ (left) and at $\beta=5.8$ (right).}
\label{3Q1}
\end{center}
\end{figure}
We study the 3Q potential $V_{\rm 3Q}$ in detail 
by investigating more than 200 patterns of the 3Q systems  
using SU(3) lattice QCD with $12^3 \times 24$ at $\beta=5.7$ and $16^3 \times 32$ 
at $\beta =5.8, 6.0$ at the quenched level. 
Here, the lattice spacing $a$ is determined so as to reproduce 
the ${\rm Q\bar Q}$ string tension $\sigma_{\rm Q\bar Q}=0.89 {\rm GeV/fm}$ at each $\beta$. 
We show in Fig.\ref{3Q1} the 3Q potential $V_{\rm 3Q}$ in SU(3) lattice QCD at $\beta$=5.7 and
$\beta$=5.8 as the function of $L_{\rm min}$, the minimal value of the total flux-tube length. 
At the large distance, where the Coulomb potential is negligible, 
we observe the linearity on $L_{\rm min}$, which is consistent with 
the expected form in Eq.(\ref{3Qp}).
At the short distance, perturbative QCD would be applicable, and therefore 
the two-body Coulomb potential in $V_{\rm 3Q}$ can be estimated as 
$V_{\rm Coul} \equiv -\frac{A_{\rm Q\bar Q}}{2}\sum_{i<j}\frac{1}{|{\bf r}_i-{\bf r}_j|}$,  
using the Coulomb coefficient $A_{\rm Q\bar Q}$ in the lattice QCD result of 
the ${\rm Q\bar Q}$ potential. (Here, the color factor $\frac12$ appears in perturbative QCD.)
To single out the confinement contribution,
we subtract this perturbative Coulomb contribution $V_{\rm Coul}$ from the 3Q potential $V_{\rm 3Q}$. 
\vspace{0.5cm}
\begin{figure}[h]
\begin{center}
%\resizebox{0.4\columnwidth}{!}{\includegraphics{3Qloop_57_cou.eps}}
%\resizebox{0.4\columnwidth}{!}{\includegraphics{3Qloop_58_cou.eps}}
\centerline{\epsfile{width=0.4\columnwidth,file=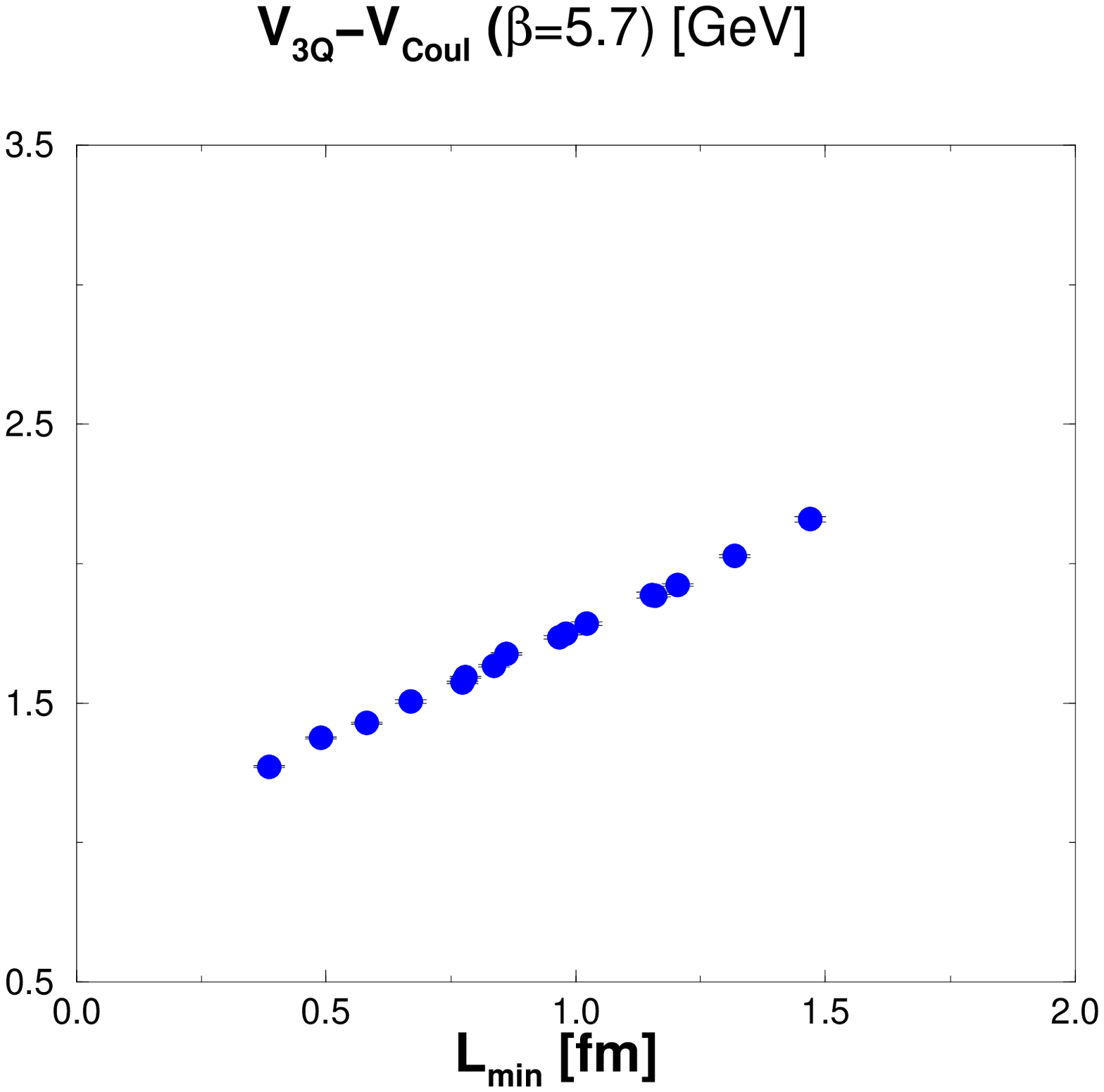}
\epsfile{width=0.4\columnwidth,file=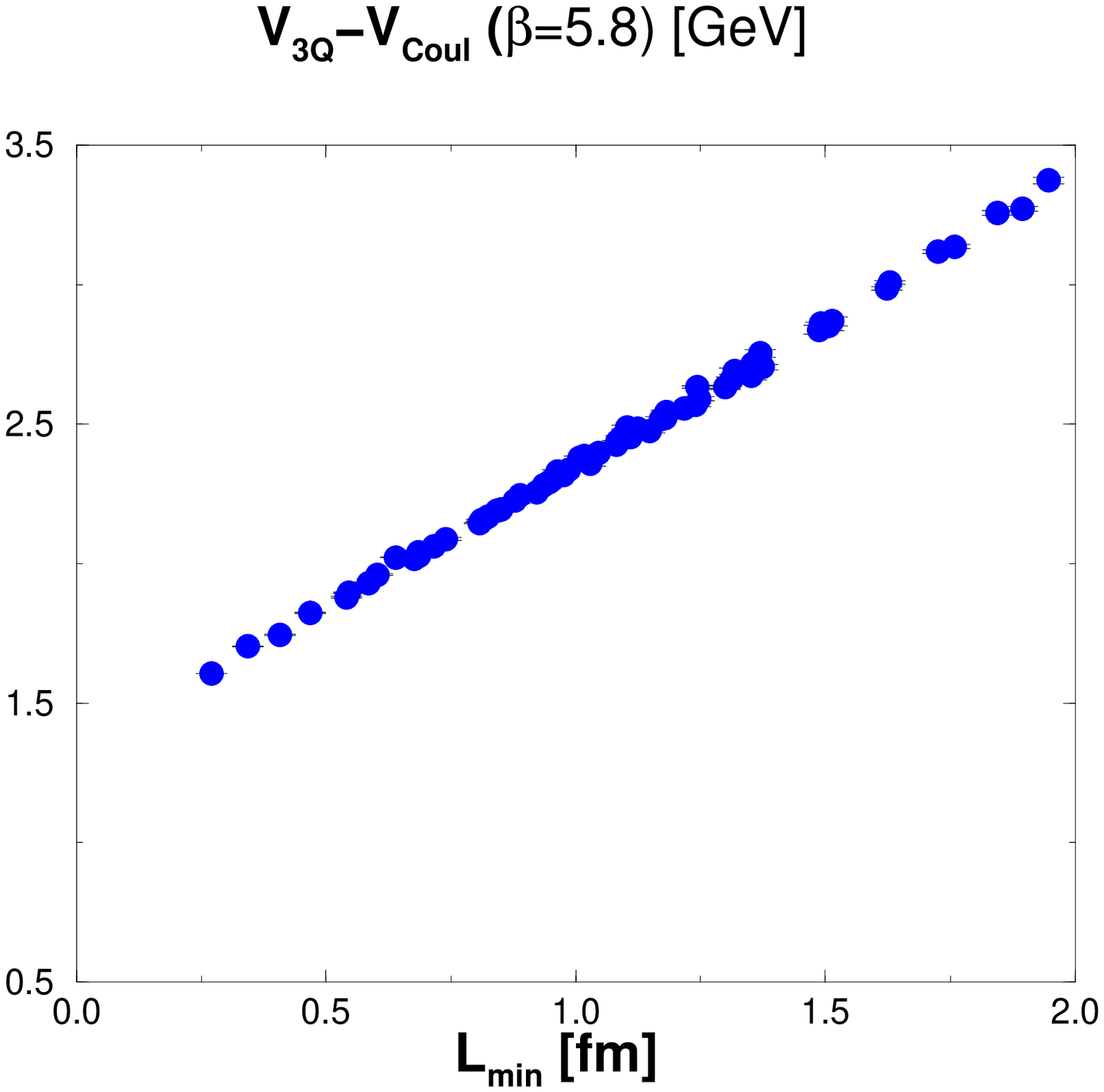}}
\caption{The SU(3) lattice QCD result for $V_{\rm 3Q}-V_{\rm Coul}$ as the function of $L_{\rm min}$, 
the minimal value of the total flux-tube length, at $\beta=5.7$ (left) and at $\beta=5.8$ (right).}
\label{3Q2}
\end{center}
\end{figure}
The results are shown in Fig.\ref{3Q2}.
The clear linearity on $L_{\rm min}$ in the whole region observed in Fig.\ref{3Q2} 
means that the 3Q potential $V_{\rm 3Q}$ can be well reproduced 
by a sum of the the perturbative Coulomb contribution as $V_{\rm Coul}$ and 
the linear confinement term proportional to $L_{\rm min}$ in Eq.(\ref{3Qp}).

Now, we consider the best fitting of the 3Q potential data with the form in Eq.(\ref{3Qp}). 
With three parameters $A_{\rm 3Q}$, $\sigma_{\rm 3Q}$ and $C_{\rm 3Q}$,
the 3Q potential data can be well reproduced within the accuracy better than a few \%.
We show the best fitting parameters of the 3Q potential $V_{\rm 3Q}$ and 
the ${\rm Q\bar Q}$ potential $V_{\rm Q\bar Q}$ in Table \ref{tab1}
at various $\beta$.
%$-$\ref{tab3} at various $\beta$.

First, we focus on the properties at the short distance, 
where the nonperturbative confinement force is almost negligible. 
We find $A_{\rm 3Q} \simeq \frac{1}{2}A_{\rm Q\bar{Q}}$ for the Coulomb coefficient, and  
find the perturbative color factor $\frac{1}{2}$ originating from the OGE contribution.

Second, let us consider the constant term. 
The constant terms $C_{\rm 3Q}$ and $C_{\rm Q\bar Q}$ behave as $\frac{1}{a}$, and 
diverge in the continuum limit $a \rightarrow 0$, so that the constant term is not physical quantity.
In the limit of $r \rightarrow 0$ in the ${\rm Q\bar Q}$ system, 
the Wilson loop becomes trivial as $W_{\rm Q\bar Q}(r=0) =1$ 
and then $V_{\rm Q \bar Q}(r=0)=0$ holds, 
while one gets $V_{\rm Q \bar Q}(r\simeq 0) \simeq - \frac{A_{\rm
Q\bar Q}}{\omega a} + C_{\rm Q\bar Q}$ with $0<\omega <1$\cite{TMNS01,SMNT01}
in the lattice regularization. 
Then, we find $C_{\rm Q\bar Q} \simeq \frac{A_{\rm Q\bar Q}}{\omega a}$.
From the similar argument on the diquark limit, 
we find also $C_{\rm 3Q}-C_{\rm Q\bar Q} \simeq \frac{A_{\rm
3Q}}{\omega a}$ \cite{TMNS01,SMNT01}.
In this way, the constant term reflects the Coulomb singularity, and is 
kept to be finite in lattice QCD, because of the lattice reguralization\cite{TMNS01,SMNT01}.  
Physically, these constant terms $C_{\rm 3Q}$ and $C_{\rm Q\bar Q}$ represent 
the pairing energy at the limit when the quarks gathers at one spatial point.
From the above two relations and  $A_{\rm 3Q} \simeq \frac{1}{2}A_{\rm Q\bar{Q}}$, 
we obtain $C_{\rm 3Q} \simeq \frac{3}{2} C_{\rm Q\bar Q}$, which seems
to hold as shown in Table 1.
%$-$3. 

%In the 3Q system, there are three pairs among three quarks with relative color
%factor 1/2 (3$\times$1/2=3/2).
%In the ${\rm Q\bar{Q}}$ system, there is only one quark
%pair. This causes the ratio of the constant terms.

Finally, we consider the infrared nonperturbative properties of the inter-quark potential.
The universality of the string tension is observed as $\sigma_{\rm 3Q}\simeq\sigma_{\rm Q\bar{Q}}$. 
In fact, the strength of the confinement force is universal for the ${\rm Q\bar{Q}}$ system and the 3Q system, 
which supports the color flux-tube picture for hadrons \cite{CI86,SST95,KS75,BPV95,RS91}.

\section{Summary and Concluding Remarks}
To bridge the gap between QCD and hadron physics, we have studied 
the 3Q potential responsible for the baryon properties.
We have calculated the ground-state 3Q potential for more than 200 patterns of the 3Q systems  
using SU(3) lattice QCD at $\beta=5.7, 5.8$ and $6.0$ at the quenched level.
In this calculation, we have adopted the smearing technique
which enhances the ground-state overlap and reduces the excited-state
contaminations like the flux vibrational modes.
Within the accuracy better than a few \%, the lattice QCD data for the 3Q potential 
have been well reproduced by a sum of a constant, the two-body Coulomb term,
and the three-body linear confinement term proportional to $L_{\rm min}$, 
the minimal value of the total flux-tube length. 
We have observed the perturbative OGE result as $A_{\rm 3Q}\simeq\frac{1}{2}A_{\rm Q\bar{Q}}$, and
the universality of the string tension as $\sigma_{\rm 3Q}\simeq \sigma_{\rm Q\bar{\rm Q}}$.

\vspace{0.5cm}
%\begin{theacknowledgments}
We acknowledge Profs. Il-Tong Cheon and Su Hong Lee for their 
warm hospitality at Yonsei University. 
The lattice QCD simulations have been performed on 
NEC-SX4 at Osaka University and HITACHI-SR8000 at KEK.
%\end{theacknowledgments}

\vspace{1cm}
\begin{table}[h]
\newcommand{\m}{\hphantom{$-$}}
\newcommand{\s}{\hphantom{1}}
\newcommand{\cc}[1]{\multicolumn{1}{c}{#1}}
\renewcommand{\tabcolsep}{2pc} % enlarge column spacing
\renewcommand{\arraystretch}{1.2} % enlarge line spacing
\begin{tabular}{@{}llll} \hline \hline
                   & \cc{$\sigma$\ (GeV/fm)} & \cc{$A$}      
& \cc{$C$\ (lattice unit)}     \\ \hline
${\rm 3Q}\ (\beta =5.7)$       & $0.832(15)$  & $0.1331(\s 66)$  & $0.9182(213)$  \\ 
%${\rm 3Q}$       & $0.1524(28)$  & $0.1331( 66)$  & $0.9182(213)$  \\ 
${\rm Q\bar{Q}}\ (\beta =5.7)$ & $0.890 (25)$  & $0.2793(116)$ & $0.6203(161)$ \\ 
%${\rm Q\bar{Q}}$ & $0.1629(47)$  & $0.2793(116)$ & $0.6203(161)$ \\ 
\hline
${\rm 3Q}\ (\beta =5.8)$       & $0.818(\s 6)$  & $0.1304(\s 17)$  & $0.9326(\s 53)$  \\ 
%${\rm 3Q}$       & $0.0999( 7)$  & $0.1304( 17)$  & $0.9326( 53)$  \\ 
${\rm Q\bar{Q}}\ (\beta =5.8)$ & $0.890 (21)$  & $0.2580(159)$  & $0.6081(182)$ \\ 
%${\rm Q\bar{Q}}$ & $0.1087(26)$  & $0.2580(159)$  & $0.6081(182)$ \\ 
\hline
${\rm 3Q}\ (\beta =6.0)$       & $0.811(\s 7)$  & $0.1363(\s 11)$  & $0.9590(\s 35)$  \\ 
%${\rm 3Q}$       & $0.0461( 4)$  & $0.1363( 11)$  & $0.9590( 35)$  \\ 
${\rm Q\bar{Q}}\ (\beta =6.0)$ & $0.890 (12)$  & $0.2768(\s 24)$  & $0.6374(\s 30)$ \\ 
%${\rm Q\bar{Q}}$ & $0.0506( 7)$  & $0.2768(\ 24)$  & $0.6374( 30)$ \\ 
\hline \hline
%\hline \hline
\end{tabular}\\[2pt]
\caption{The best fitting parameters for the 3Q potential $V_{\rm 3Q}$
at $\beta$=5.7, 5.8 and 6.0.}
\label{tab1}
\end{table}

\newpage
\small
\begin{flushleft}
{\rm \bf Table 2.} \ \ 
Partial data for the 3Q potential in SU(3) lattice QCD at $\beta$=5.8. 
$(i,j,k)$ labels the 3Q system where the three quarks 
put on $(i,0,0)$, $(0,j,0)$ and $(0,0,k)$ in ${\bf R}^3$ in the lattice unit.
\end{flushleft}
\begin{minipage}{0.5\columnwidth}
%\vspace{0.2cm}
%\begin{center}
\newcommand{\m}{\hphantom{$-$}}
\newcommand{\cc}[1]{\multicolumn{1}{c}{#1}}
\begin{tabular}{c c c c} \hline\hline
$(i, j, k)$ & $V_{\rm 3Q}^{\rm latt}$
& $V_{\rm 3Q}^{\rm fit}$ & $V^{\rm latt}_{\rm 3Q}-V^{\rm fit}_{\rm 3Q}$ \\
\hline
$(1, 1, 1 )$ &     0.9144 &     0.9008 &    \m0.0136            \\
$(1, 1, 2 )$ &     1.0657 &     1.0582 &    \m0.0075            \\
$(1, 1, 3 )$ &     1.1962 &     1.1883 &    \m0.0079            \\
$(1, 1, 4 )$ &     1.3049 &     1.3054 &    $-$0.0005           \\
$(1, 1, 5 )$ &     1.4134 &     1.4162 &    $-$0.0029           \\
$(1, 1, 6 )$ &     1.5339 &     1.5236 &     \m0.0103           \\
$(1, 2, 2 )$ &     1.1866 &     1.1877 &    $-$0.0011           \\
$(1, 2, 3 )$ &     1.3062 &     1.3076 &    $-$0.0014           \\
$(1, 2, 4 )$ &     1.4153 &     1.4203 &    $-$0.0050           \\
$(1, 2, 5 )$ &     1.5259 &     1.5287 &    $-$0.0029           \\
$(1, 2, 6 )$ &     1.6360 &     1.6347 &     \m0.0013           \\
$(1, 3, 3 )$ &     1.4178 &     1.4209 &    $-$0.0031           \\
$(1, 3, 4 )$ &     1.5216 &     1.5296 &    $-$0.0080           \\
$(1, 3, 5 )$ &     1.6276 &     1.6357 &    $-$0.0081           \\
$(1, 3, 6 )$ &     1.7296 &     1.7401 &    $-$0.0105           \\
$(1, 4, 4 )$ &     1.6291 &     1.6356 &    $-$0.0064           \\
$(1, 4, 5 )$ &     1.7190 &     1.7396 &    $-$0.0206           \\
$(1, 4, 6 )$ &     1.8203 &     1.8425 &    $-$0.0222           \\
$(2, 2, 2 )$ &     1.2793 &     1.2838 &    $-$0.0046           \\
$(2, 2, 3 )$ &     1.3815 &     1.3903 &    $-$0.0089           \\
$(2, 2, 4 )$ &     1.4904 &     1.4969 &    $-$0.0064           \\
$(2, 3, 3 )$ &     1.4751 &     1.4880 &    $-$0.0129           \\
$(2, 3, 5 )$ &     1.6814 &     1.6912 &    $-$0.0098           \\
%$(2, 3, 6 )$ &     1.7847 &     1.7931 &    $-$0.0084           \\
%$(2, 4, 4 )$ &     1.6800 &     1.6869 &    $-$0.0069           \\
%$(2, 4, 5 )$ &     1.7669 &     1.7862 &    $-$0.0193           \\
%$(3, 3, 3 )$ &     1.5562 &     1.5746 &    $-$0.0185           \\
%$(3, 3, 4 )$ &     1.6475 &     1.6692 &    $-$0.0217           \\
%$(3, 4, 6 )$ &     1.9425 &     1.9496 &    $-$0.0071           \\
%$(3, 5, 5 )$ &     1.9322 &     1.9446 &    $-$0.0124           \\
\hline\hline
\end{tabular}\\[2pt]
%\end{center}
\end{minipage}
\begin{minipage}{0.5\columnwidth}
%\vspace{0.6cm}
%\begin{center}
\newcommand{\m}{\hphantom{$-$}}
\newcommand{\cc}[1]{\multicolumn{1}{c}{#1}}
\begin{tabular}{c c c c} \hline\hline
$(i, j, k)$ & $V_{\rm 3Q}^{\rm latt}$
& $V_{\rm 3Q}^{\rm fit}$ & $V^{\rm latt}_{\rm 3Q}-V^{\rm fit}_{\rm 3Q}$ \\
\hline
$(4, 5, 6 )$ &     2.1263 &     2.1094 &     \m0.0169           \\
$(4, 6, 6 )$ &     2.2264 &     2.1979 &     \m0.0285           \\
$(5, 6, 6 )$ &     2.3105 &     2.2733 &     \m0.0372           \\
$(0, 1, 1 )$ &     0.7701 &     0.7727 &    $-$0.0027           \\
$(0, 1, 2 )$ &     0.9659 &     0.9694 &    $-$0.0035           \\
$(0, 1, 3 )$ &     1.1107 &     1.1071 &     \m0.0037           \\
$(0, 1, 4 )$ &     1.2323 &     1.2268 &     \m0.0055           \\
$(0, 1, 5 )$ &     1.3339 &     1.3388 &    $-$0.0049           \\
$(0, 1, 6 )$ &     1.4566 &     1.4469 &     \m0.0097           \\
$(0, 1, 7 )$ &     1.5730 &     1.5527 &     \m0.0203           \\
$(0, 1, 8 )$ &     1.6779 &     1.6570 &     \m0.0209           \\
$(0, 2, 2 )$ &     1.1375 &     1.1422 &    $-$0.0047           \\
$(0, 2, 3 )$ &     1.2660 &     1.2711 &    $-$0.0050           \\
$(0, 2, 4 )$ &     1.3815 &     1.3871 &    $-$0.0056           \\
$(0, 2, 5 )$ &     1.4928 &     1.4972 &    $-$0.0043           \\
$(0, 2, 6 )$ &     1.6069 &     1.6041 &     \m0.0028           \\
$(0, 2, 8 )$ &     1.8285 &     1.8128 &     \m0.0157           \\
$(0, 3, 3 )$ &     1.3893 &     1.3940 &    $-$0.0047           \\
$(0, 3, 4 )$ &     1.5007 &     1.5066 &    $-$0.0059           \\
$(0, 3, 5 )$ &     1.6135 &     1.6145 &    $-$0.0010           \\
$(0, 4, 4 )$ &     1.6079 &     1.6165 &    $-$0.0086           \\
$(0, 4, 5 )$ &     1.7085 &     1.7226 &    $-$0.0141           \\
$(0, 4, 6 )$ &     1.8263 &     1.8267 &    $-$0.0004           \\
%$(0, 4, 7 )$ &     1.9328 &     1.9297 &     \m0.0031           \\
%$(0, 4, 8 )$ &     2.0399 &     2.0319 &     \m0.0080           \\
%$(0, 5, 5 )$ &     1.7994 &     1.8271 &    $-$0.0277           \\
%$(0, 5, 6 )$ &     1.9205 &     1.9300 &    $-$0.0096           \\
%$(0, 6, 6 )$ &     2.0302 &     2.0319 &    $-$0.0018           \\
%$(0, 6, 7 )$ &     2.1390 &     2.1330 &     \m0.0060           \\
%$(0, 6, 8 )$ &     2.2371 &     2.2337 &     \m0.0035           \\
\hline\hline
\end{tabular}\\[2pt]
%\end{center}
\end{minipage}

\end{document}